\newcounter{myctr}
\def\myitem{\refstepcounter{myctr}\bibfont\noindent\ifnum\themyctr>9\else\phantom{0}\fi\hangindent17pt\themyctr.\enskip}

\documentclass{ijqi-2e/ws-ijqi}

\usepackage{hyperref}
\usepackage[super,sort,compress]{cite}
\usepackage[dvipsnames]{xcolor}
\hypersetup{colorlinks=true, allcolors=blue, pdfusetitle=true}

\usepackage{tikz}
\usetikzlibrary{zx-calculus}
\usetikzlibrary{quantikz2}	

\usepackage{braket}

\makeatletter
\renewcommand\appendix{\refstepcounter{appendix}
	\setcounter{section}{0}%
        \setcounter{lemma}{0}
        \setcounter{theorem}{0}
	\setcounter{definition}{0}
        \setcounter{corollary}{0}
        \setcounter{conjecture}{0}        
	\setcounter{equation}{0}
	\@addtoreset{equation}{section}
	\renewcommand\theequation{\Alph{section}.\arabic{equation}}
	\renewcommand\sectionmark[1]{\markright{\thesection\ ##1:}} 
	\@ifstar\appendixstar\appendixnostar}
\newcommand\appendixnostar{
	\renewcommand\thesection{\appendixname\ \Alph{section}}}
\newcommand\appendixstar{
	\renewcommand\thesection{\appendixname}}
\makeatother

\begin{document}

\title{\uppercase{Simon algorithm in measurement-based quantum computing}}
\author{M. SCHWETZ and R. M. NOACK}
\address{Fachbereich Physik, Philipps Universität Marburg, Renthof 6\\
35037 Marburg, Germany\\
maximilian.schwetz@physik.uni-marburg.de}
\date{\today}

\maketitle


\begin{abstract}

Simon's hidden subgroup algorithm was the first quantum algorithm to
prove the superiority of quantum computing over classical computing in
terms of complexity.
Measurement-based quantum computing (MBQC) is a formulation of quantum
computing that, while equivalent in terms of
computational power, can be advantageous in experiments and
in displaying the core mechanics of quantum algorithms. 
We present a reformulation of the Simon algorithm into the language of
MBQC---in detail for two qubits and schematically for $n$ qubits.
We utilize the framework of ZX-calculus, a graphical tensor description of
quantum states and operators, to translate the circuit description of the
algorithm into a form concordant with MBQC. 
The result for the two-qubit Simon
algorithm is a ten-qubit cluster state on which single-qubit
measurements suffice to extract the desired information.
Additionally, we show that the $n$-qubit version of the Simon algorithm
can be formulated in MBQC as
cluster state graph with $2n$ nodes and $n^2$ edges.
This is an example of the MBQC formulation of a quantum algorithm that
is exponentially faster than its classical counterpart.
As such, this formulation should aid in
understanding the core mechanics of such an
established algorithm and could serve
as a blueprint for experimental implementation.

\end{abstract}


\markboth{Schwetz and Noack}
{Three-qubit Deutsch-Jozsa in MBQC}

\section{Introduction}

Measurement-based quantum computing (MBQC) is an alternate
approach to quantum computing that is conceptually different than
the canonically used circuit model \cite{MBQC-Review}.
In contrast to the circuit model, in which
entangling multi-qubit gates
effectuate the computing power of quantum mechanics,
MBQC pre-encodes the entanglement in a
highly entangled multi-qubit initial state that is prepared in
advance. 
Typically, this resource state is a cluster state, an eigenstate of the
$ZXZ$-stabilizer that
is the ground state of lattices with
Ising interaction \cite{Cluster-state}.
Other states, such as the two-dimensional AKLT state with spin greater
than $3/2$,
are also possible, but the information processing protocol will deviate from
the canonical approach to MBQC \cite{AKLT-MBQC-1, AKLT-MBQC-2}.
Single-qubit measurements on the entangled initial state
are then sufficient to implement quantum computing in full generality
\cite{onewayQC, Cluster_QC, MBQC-Review}.
The MBQC model can lead to alternate experimental
implementations of quantum algorithms that can, in some cases, be more
flexible and efficient than circuit-based implementations.
In addition, it is useful for highlighting and understanding certain fundamental
aspects of quantum computing that are not equally prominent in
circuit-based quantum computing.

From an experimental point of view, it is convenient that, in MBQC,
the quantum-specific part in the form of entanglement is
carried out at the beginning of the process.
This preparation of the initial cluster state could be outsourced
to a device whose sole purpose is generating entanglement.
The prepared state would then be handed over to a device that 
need only carry out one-qubit measurements.
The latter are---in comparison with entangling multi-qubit gates in
a circuit-based setup---easy to perform.
The composite device would, in effect, function as a universal quantum
computer.
On a fundamental level,
MBQC leads to a compelling alternate picture of quantum computing:
one reason for the advantage
of quantum computing over classical computing can,
among other properties, be found in the
necessity to generate a multi-qubit entangled state.
Hence, the creation of such a state is generally difficult because
it cannot be carried out by any classical device.
The measurements carried out in the
subsequent phase of MBQC can be interpreted as
classical tools that utilize the previously
created resource of quantum entanglement for calculational 
purposes.
The pedagogical appeal of this framework is that
the ``quantumness'' and the classical operations are clearly separated.
This picture can potentially provide a better understanding of where
the advantage over classical computing actually lies.

In other contexts, the circuit model might
nevertheless
be the more suitable framework to utilize.
The set of tools for information processing in the circuit model is
analogous to the logic-gate model of classical computing
in which elementary logic gates are used as building blocks to compose
a complex algorithm.
The decomposition into a small set of fundamental elements
as well as the resulting linear flow of information processing from left
to right in a diagram reminiscent of the canonical model of classical
computing.
In MBQC, the flow of information processing during a computation is
much less evident due to the multi-dimensional architecture of a graph
state.
In addition, in the quantum circuit model,
the actual effect of one gate on a set of qubits can be
denoted much more compactly and understandably than
its analog in MBQC, the effect of a
single-qubit measurement on a highly entangled multi-partite state.

In MBQC, the key element for information processing is a
sequence of measurements on the qubits of a cluster state.
The measurements are carried out along selected axes to realize
a specific algorithm.
Models such as the \emph{measurement calculus} \cite{Calculus}
or the \emph{monad} \cite{Monads} description
have been formulated to
try to encompass
the complexity of MBQC.
In this paper,
we will utilize the graphical notation of the \emph{ZX-calculus}, which
is able to represent graph states as well as graph-like operators in a
unified way \cite{ZX-calculus}.
This graphical calculus was introduced
for carrying out derivations in multi-qubit quantum computation and
information.
On a computational level, the diagrams are tensor-network descriptions of
both quantum states and quantum operators; in fact, the notation does not
distinguish between the two.
The ZX-calculus is also naturally suited to describe cluster states in
arbitrary topologies and has a primitive notation to denote
projective, single-qubit measurements.
Therefore, it is a framework that is well-suited for describing MBQC in an
intuitive manner.

The set of calculation rules that can be applied within
ZX-calculus are not only
used for transforming ZX-diagrams into one another, but are also capable
of translating an algorithm formulated in terms of
quantum circuits to the language of MBQC \cite{ZX-simplification}.
The aforementioned rules are equations that tell the user how to transform
ZX-diagrams.
On a low level, they simply represent tensor equations.
They range from single-node identities
to large-scale node and edge manipulations in the graph of a ZX-diagram.
One application for these rules is to simplify diagrams and thus to
optimize quantum circuits \cite{ZX-simplification}.
Another application of the ZX-calculus is 
to translate quantum circuits into the language
of ZX-diagrams; we will do this here.
Subsequently, we will use the rules mentioned above to
transform these diagrams into a specific format
to reformulate the oracle of the Simon algorithm as
a measurement-based quantum calculation.
This work directly builds on
previous work of the authors in which the three-qubit Deutsch-Josza
algorithm in MBQC is derived.\cite{3-qubit-DJ-MBQC}

The algorithm proposed by Daniel Simon in 1993 and published in 1994 was
the first quantum algorithm that was proven to be exponentially faster than
any classical computation on the same problem \cite{Simon}.
The algorithm solves a ``promise problem'' in which a
``black-box'' oracle executes 
a function with a property that is not known initially but is promised
to lie within a given set of properties.
The specific proposal by Simon was to
find the period of a function, that is, the numerical distance between two
inputs that trigger the same output. The algorithm was later generalized
to solve a general \emph{hidden subgroup} problem on arbitrary groups
\cite{Hidden-subgroup, Hidden-subgroup-non-abelian}.
Even though Simon did not fully realize the significance of his
development immediately, the Simon algorithm
subsequently inspired Peter Shor to apply
a similar theoretical idea
to the problmes of factoring and discrete logarithms \cite{Simon-Interview}.
Eventually, this lead to the development Shor's famous algorithm
for prime factorization, a lighthouse
in the realm of quantum algorithms
\cite{Shor}.

Our goal will be to derive a measurement-based formulation
of the Simon algorithm in a systematic way using the ZX-calculus.
We will start by introducing the Simon algorithm in the canonical
language of circuit-based quantum computing in Sec.~\ref{sec:simon}.
In a deeper analysis, we will investigate the forms of the possible
oracles and will formulate a deterministic algorithm to translate a given
periodic function to a circuit for the oracle in Sec.~\ref{sec:oracle}.
Sec.~\ref{sec:MBQC} will serve to reformulate
the oracle circuits in terms of MBQC and will use the two-qubit
version of the algorithm as an example.
The result will be a ZX-diagram that incorporates both the form
of the required cluster state as well as a recipe for the single-qubit
measurements that realizes the Simon algorithm in MBQC.
After having illustrated the concepts and procedure for the two-qubit
case, in Sec.~\ref{sec:nbit} we will go on to describe the topology of
resource cluster states for $n$ qubits.
We will observe that the algorithm gains complexity in the sense of
the increase in the number of cross sections within the required
graph state pictured in two dimensions.
Finally, in Sec.~\ref{sec:conclusion}, we will discuss
the key aspects of this work and give an outlook for potential future
work.

\section{Simon algorithm}
\label{sec:simon}

A function $f$ is said to have a period $s$ if all elements of
its  domain that are separated by $s$  yield the same image under $f$.
Consider a function $f: \mathbb{S}^n \rightarrow \mathbb{S}^n$, where
$\mathbb{S} = \{0, 1\}$, which
maps from $n$-bit binary numbers to $n$-bit binary numbers.
Its period $s$ is defined as
\begin{equation}
	\exists! \; s \in \mathbb{S}^n: \; \forall \; a,
	b \in \mathbb{S}^n, \  a \ne b: \, f(a) =
	f(b) \; \Leftrightarrow \; a = b \oplus s \, , 
\end{equation}
where $\oplus$ is bitwise addition modulo 2.
From now on we will use the terms \emph{periodic} and \emph{period}
to refer only to a function that 
fulfills the aforementioned definition of periodicity modulo 2 and not
other definitions of periodicity.
When $s \ne 0$, the function is \emph{two-to-one}.
and when $s = 0$, the function is \emph{bijective}.
Given such a periodic function, the aim of the Simon algorithm is to
determine the period $s$ of $f$.
We assume that the function is only accessible as a black box
(oracle), which can act on a (quantum) input, but that no
information about the internal properties of the black box is available.

The Simon algorithm works by applying the function $f$ to an $n$-qubit
quantum state that is a superposition of all possible inputs.
The output of $f$ is stored on $n$ auxiliary qubits, and
each pair of input basis states that are $s$ apart are
``marked'' by the same basis state as on the auxiliaries.
This scheme generates
entanglement that groups together
pairs of basis states that are a period $s$ apart.
A quantum-interference-based
filter subsequently is used to
filter out a bit string $t$ that fulfills the linear
equation $s \oplus t = 0$.
After the procedure is repeated sufficiently many times to 
obtain a sufficiently large set of linearly independent values of $t$,
a classical solver for systems of linear equations can be used to
solve for the period $s$.

\begin{figure}
\begin{center}
\begin{quantikz}
	\lstick{$\ket{+^{\otimes n}}$} & \gate[2][2cm]{\text{oracle}}\gateinput{$t$}\gateoutput{$t$} & \meter{x} \\
	\lstick{$\ket{0^{\otimes n}}$} & \gateinput{$u$} \gateoutput{$u\oplus f(t)$} & \qw
\end{quantikz}
\caption{Simon algorithm in the quantum circuit framework.}
\label{fig:circuit}
\end{center}
\end{figure}
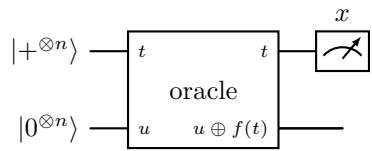

We start with $n$ working qubits, where $n$ is equal to the dimension
of the function domain and image, each prepared to be in the state
$\ket{+} \simeq \ket{0}+\ket{1}$.
(Note that we omit normalization here and throughout this paper.)
In addition to the working qubits,
we will need
$n$ auxiliary qubits, initially prepared to be in the
state $\ket{0}$.
The full Simon algorithm, including the initial states and the
measurement of the final state, is depicted schematically in
Fig.~\ref{fig:circuit}.

%
We assume that the oracle applies
the function $f$ to a
register of $n$ working qubits,
putting the result in the register of $n$
auxiliary qubits.
In doing so, we additionally
assume that the oracle takes into account the output
value of the function $f$ and maps a general basis
state of the full system (working and auxiliary register)
$\ket{t}\ket{x}$ as 
$\ket{t}\ket{x} \rightarrow \ket{t}\ket{x \oplus f(t)}$.
Since all qubits of
the auxiliary register are prepared in the $\ket{0}$-state,
the effect of the oracle is to map
$\ket{t}\ket{0^n} \rightarrow \ket{t}\ket{f(t)}$,
and we end up with the image under $f$ of the working
register in the auxiliary register.
Thus, after the application of the oracle, the system is in the entangled state
\begin{equation}
	\ket{\psi} = \sum_{t\in\{0,1\}^n} \ket{t}\ket{f(t)} \; . \label{eq:after_oracle}
\end{equation}
In effect, basis states $\ket{t}$ and
$\ket{t'}$ for which  $t' = t \oplus s$ in the working
register are ``marked'' by the same
basis state of the auxiliary register,
$\ket{f(t)} = \ket{f(t')}$.

The final step of the quantum algorithm as depicted in
Fig.~\ref{fig:circuit} is the measurement of the
qubits of the working register
in their respective
$x$-bases; the possible measurement outcomes 
will lie in $\{+, -\}^n$.
However, instead of carrying out the measurements in the $x$-basis, we could
apply an $n$-qubit Hadamard gate $H^{\oplus n}$ to the working
register and then carry out the measurements in the computational $z$-basis.
After the Hadamard gate, the total state in eq.~(\ref{eq:after_oracle}) transforms
to
\begin{equation}
\begin{split}
	\ket{\psi'} &= \sum_{r\in\{0,1\}^n} \sum_{t\in\{0,1\}^n} (-1)^{r\cdot t} \ket{r}\ket{f(t)} \\
		&= \sum_{r\in\{0,1\}^n} \sum_{b \in \text{Im}(f)} \sum_{\substack{a\\f(a) = b}} (-1)^{r \cdot a} \ket{b} \; . \label{eq:after_hadamard}
\end{split}
\end{equation}
%
If $s = 0^n$, $f$ is a \emph{one-to-one} function,
and $\ket{\psi'}$ simplifies to 
\begin{equation}
\begin{split}
	\ket{\psi'_{s=0}} = \sum_{r\in\{0,1\}^n} \sum_{t\in\{0,1\}^n} (-1)^{r\cdot t} \ket{r} \; . \label{eq:s0_after_hadamard}
\end{split}
\end{equation}
If we now perform a measurement on the working
qubits---which are in state
(\ref{eq:s0_after_hadamard})---in the computational basis, all 
measurement outcomes have the same probability.
This also holds for the original case in which the $n$-qubit Hadamard gate is
not present, and the measurement is carried out in the $x$-basis.
As $s=0^n$ in this case, all outcomes $m$ satisfy $s \oplus m = 0$.

For the case of a non-vanishing period $s \ne 0^n$, $f$
is a \emph{two-to-one} function, and
eq.~(\ref{eq:after_hadamard}) can be rewritten as 
\begin{equation}
	\ket{\psi'} = \sum_{r\in\{0,1\}^n} \sum_{b \in \text{Im}(f)} \Big[(-1)^{r \cdot a_1} + (-1)^{r \cdot a_2}\Big] \ket{r} \ket{b} \; , \label{eq:s1_after_hadamard}
\end{equation}
where $f(a_1) = f(a_2) = b$.
Using that $a_2 = a_1 \oplus
s$, we obtain
\begin{equation}
\begin{split}
	\ket{\psi'} &= \sum_{r\in\{0,1\}^n} \sum_{b \in \text{Im}(f)} \Big[(-1)^{r \cdot a_1} + (-1)^{r \cdot a_1 \oplus s}\Big] \ket{r} \ket{b} \\
		&= \sum_{r\in\{0,1\}^n} \sum_{b \in \text{Im}(f)} \Big\{(-1)^{r \cdot a_1} \big[1 + (-1)^{r \cdot s}\big]\Big\} \ket{r} \ket{b}\; .
\end{split}
\label{eq:s1_after_hadamard_simplified}
\end{equation}
The term in square brackets in the last line of
eq.~(\ref{eq:s1_after_hadamard_simplified}) will vanish if and only if
$r \cdot s = 1$ and will be
non-zero only when $r \cdot s = 0$.
Thus, measuring the working qubits in the computational basis will
generate an outcome $m$ that satisfies $m \cdot s = 0$.
This result also holds for the circuit without the $n$-qubit
Hadamard gate as long as the measurement is carried out
in the $x$-basis.

After a single measurement of the working qubits, we obtain one binary number
$m$ that satisfies $m \cdot s = 0$, but not $s$ itself.
In order to determine $s$, the quantum part of the
algorithm must be
executed repeatedly until $n-1$ linearly independent outcomes $m$
are obtained.
The resulting set of $n-1$ linear equations can be solved using a
classical algorithm to obtain $s$, provided that $s \ne 0^n$.
If $s = 0^n$, the result is indeterminate, as any $m$ will satisfy  $m \cdot s = 0$.
To make sure that $s \ne 0^n$, we can simply apply $f$ to two
different inputs $a_1$ and $a_2$ that satisfy $a_2 = a_1 \oplus s$,
where $s$ is the solution of the set of linear equations.
If $f(a_1) = f(a_2)$, we know that $f$
is a \emph{two-to-one} function for which we have
obtained the correct period $s$.
If, on the other hand, $f(a_1) \ne f(a_2)$,
the function $f$ must be \emph{one-to-one} with $s = 0^n$.

\section{Oracle}
\label{sec:oracle}

\subsection{Oracle design}

Up to 
now, we have conceptualized the oracle as a black box that
applies a particular instance $f$ of a class of possible
oracle functions to an arbitrary input state consisting of $n$ working qubits
and $n$ auxiliary qubits.
Mathematically, the operation that is carried out is to add the image under $f$
of the state of the working qubits to the state of the auxiliary
qubits, as has already been visualized in Fig.~\ref{fig:circuit}.
For practical application, however, the oracle must have a physical
implementation that depends on and implements a particular specific
instance $f$.
We now consider how to carry out such an implementation in a quantum
circuit picture with the aim of subsequently reformulating the
implementation in MBQC.

Since the function of the oracle is to add bits (modulo 2) to the auxiliary
qubits conditionally depending on the state of the working qubits, it
is natural to use CNOT gates to construct the circuit implementing
$f$.
A CNOT gate will, for a two-qubit basis state,
add a bit (modulo 2) to the state of the second
qubit if and only if the first qubit is in the state
$\ket{1}$. Mathematically, the gate can be written as
\begin{equation}
	\text{CNOT} = \ket{00}\bra{00} + \ket{01}\bra{01} + \ket{10}\bra{11} + \ket{11}\bra{10} \; .
\end{equation}
Since the oracle can include information flow from all working to
all auxiliary qubits, a circuit formulation should also be able to
include any CNOT gate that has one of the working qubits as its
control and one of the auxiliary qubits as its target.

Normally, the CNOT gate is only activated when the first (control) qubit
is 1.
However, here it is convenient to be able to activate the CNOT using a
0 as the control condition.
We can do this by adding an $X$ gate to the auxiliary qubit.
Since the function of an $X$ gate is to swap each bit in the qubit
basis, following a CNOT with an $X$ gate implements a control gate in
which the bit is flipped in the second qubit's basis if and only if
the basis state of the first qubit is $\ket{0}$.
As there is freedom to displace the $X$ gates to the right in the quantum
circuit, we group all $X$ gates as the last possible operation of the
oracle.

Note that the $X$ gates described above are actually not required for
the algorithm.
As is evident in Fig.~\ref{fig:circuit}, adding single-qubit gates
to the auxiliary register does not affect the
outcome of the measurement on the working register.
Indeed, the essential ingredient for finding the period is the entanglement
between working and auxiliary qubits. 
Nevertheless, a case may be made for retaining the $X$ gates within our
theoretical treatment.
The primary reason is that we need to check that the found period $s$ is correct in
the classical post-quantum processing of the algorithm (e.g., in order
to determine if $s=0$).
For a general oracle, the value of $f$ is available in the auxiliary
qubits, and we can check any value by sampling.
While a reduced oracle without $X$ gates will still be able to tell us if
the images of two samples under $f$ are the same, it will not yield the true
value of the image under $f$.
In this paper, we retain the $X$ gates, implementing the most general
version of the oracle.

\begin{figure}
	\begin{center}
	\begin{quantikz}
		\lstick{$\ket{+}$} & \ctrl{2} & \ctrl{3} & \qw & \qw & \qw & \meter{x} \\
		\lstick{$\ket{+}$} & \qw & \qw & \ctrl{1} & \ctrl{2} & \qw & \meter{x} \\
		\lstick{$\ket{0}$} & \targ{} & \qw & \targ{} & \qw & \gate{X} & \qw \\
		\lstick{$\ket{0}$} & \qw & \targ{} & \qw & \targ{} & \gate{X} & \qw
	\end{quantikz}
	\end{center}
	\caption{One particular instance of the 2-qubit oracle for Simon's algorithm along with the the initial states and measurement operations. The oracle instance shown in the figure is the one with the maximum number of gates. Other instances of $f$ imply that some of the shown gates are ``switched off''.}
	\label{fig:general-circuit}
\end{figure}
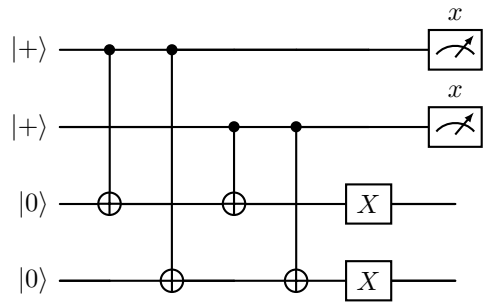

A specific set of such gates for one particular 
instance $f$ of the oracle function is depicted in
Fig.~\ref{fig:general-circuit}.
Indeed, this instance 
contains the maximum number of
possible gates for oracles in the Simon algorithm.
Other instances of $f$ would require fewer gates,
that is, some of the gates in Fig.~\ref{fig:general-circuit} would be present
and some absent.
Each unique combination from this maximum set of possible gates
realizes one particular instance of $f$.

\subsection{Generalization}

Now that we have shown that the oracle for any
$n$-bit periodic function $f$ can be realized as a network consisting
of CNOT gates between $n$ working and $n$ auxiliary qubits and $X$
gates acting on the auxiliary qubits, we want to
reformulate the network for such an oracle in MBQC.
In order to do this, we must
first gain a formal understanding of which realization 
of the function $f$ corresponds to
what combination of CNOTs and $X$ gates.
In the following, we formulate a deterministic algorithm to construct
a quantum network implementing the oracle for a given periodic
function $f$.

A general $n$-qubit function $f$ has $2^n$ possible outputs
($|\mathrm{Im}\, f| = 2^n$).
Each output $\boldsymbol{\omega}_{\boldsymbol\sigma}$ is the image of
a basis state $\boldsymbol\sigma$ of the domain.
For example, $\boldsymbol\sigma = 010$ might be a basis state and
$\boldsymbol\omega_{\boldsymbol\sigma} = 110$ its corresponding image,
so $f(010) = 110$.
We define the \emph{characteristic} $\mathcal{C}_f$ of the function
$f$ as the output string of all computational-domain basis states
\begin{equation}
	\mathcal{C}_f = \prod_{\boldsymbol\sigma\in\mathbb{Z}_2^n}
	``\omega_{\boldsymbol\sigma}"
	= \omega_{0\dotsm00} \, \omega_{0\dotsm01} \dotsm \,\omega_{1\dotsm11} \;
	, \label{eq:vals}
\end{equation}
where the double quotes ``\ldots''
indicate that the meaning of the product is to concatenate the binary
numbers.
For $n$ qubits, the characteristic of a function is a string of length
$n\,2^n$ bits.
For example, the two-to-one function $g$ that is defined by
\begin{equation}
	g(00) = 10 \, , \quad g(01) = 01 \, , \quad g(10) = 01 \, , \quad g(11) = 10
\end{equation}
would be represented by the characteristic
\begin{equation}
	\mathcal{C}_g = \underbracket{10}_{g(00)}\underbracket{01}_{g(01)}\underbracket{01}_{g(10)}\underbracket{10}_{g(11)} \; .
\end{equation}

Any classical mapping $f$ with period modulo 2 between input and output qubits can be
implemented as a quantum network consisting of CNOT gates
acting on particular pairs of input and output qubits along with $X$ gates on the
output qubits.
We will encode a configuration of these gates as a
concatenated binary string of length  $n 2^n$, as we have done for
the characteristic of the function in eq.~(\ref{eq:vals}).
We define the characteristic of a CNOT gate between input qubit $i$ and output
qubit $j$ to have ones at the bit positions at which the corresponding basis
vector is changed.
Mathematically, we define the characteristic of a CNOT gate as
\begin{equation}
	\big[\mathcal{C}_{\mathrm{CNOT}, j, k}\big]_m = \begin{cases}
	1 \, , & m = (2^n - l j - 1) n + k\, , \; \forall
	l \in \mathbb{N}, \, 0 \le l < n \\0 \, ,  & \text{otherwise}
        \end{cases} \; , \label{eq:CNOT}
\end{equation}
where $(j, k \in \mathbb{N}, \, 1 \le j, k \le n)$.
In words, we divide the characteristic into
$2^n$ sets of $n$ bits, with each set corresponding to one input basis
state.
That is, the first $n$ bits are the image of the first basis state,
the subsequent $n$ bits are the image of the second basis state, and so
on.
A CNOT gate will generate non-zero bits only in the $n$-bit sets that
correspond to basis states for which the tested control bit is $1$.
Each bit of these $n$-bit sets corresponds to one of the $n$ output
bits, so the first bit of the characteristic is the first output bit
of the first input basis state.
Within such a set, the bit corresponding to the target qubit is $1$,
and all others are $0$.

\begin{figure}\begin{center}
	\begin{quantikz}
		\lstick[2]{domain} & \ctrl{3} \wire[l][1]["1"{above,pos=0.8}]{a} & \\
		& \qw \wire[l][1]["2"{above,pos=0.8}]{a} & \\
		\lstick[2]{image} & \qw \wire[l][1]["1"{above,pos=0.8}]{a} & \\
		& \targ{} \wire[l][1]["2"{above,pos=0.8}]{a} &
	\end{quantikz} \; .
	\caption{Circuit for the CNOT gate with control bit 1 and target bit 2 for a 2-bit system.}
	\label{fig:CNOT-example}
\end{center}\end{figure}
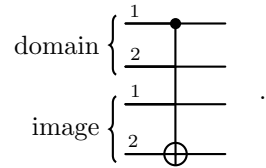

As an example, we consider a system with two domain and two image
bits.
This directly implies that the characteristic of both the function
and the gates have a length of $2 \cdot 2^2 = 8$ bits.
Furthermore, we take the CNOT gate to act between qubit 1 of the domain and
qubit 2 of the image, as depicted in the circuit of
Fig.~\ref{fig:CNOT-example}.
The characteristic of this gate, according to eq.~(\ref{eq:CNOT}), is
\begin{equation}
	\big[\mathcal{C}_{\mathrm{CNOT}, 1, 2}\big]_m = \begin{cases} 1 \, , & m \in \{5, 7\} \\0 \, , & \text{otherwise} \end{cases} \quad \Rightarrow \quad \mathcal{C}_{\mathrm{CNOT}, 1, 2} = 00000101 \; .
\end{equation}
This characteristic has the following content:
The control qubit is the first qubit.
Hence, the third and fourth set of $n = 2$ bits, corresponding to the
basis vectors $10$ and $11$, are not all zero.
Within these sets, the second bit is set to $1$
because the target qubit of the CNOT gate is the second of the image qubits.

Analogously to the CNOT gate of
eq.~(\ref{eq:CNOT}), we can 
characterize the $X$ gate on image qubit $k$ as 
\begin{equation}
	\big[\mathcal{C}_{X, k}\big]_m = \begin{cases} 1 \, , & m = l n + k \, , \; \forall l \in \mathbb{N}, \, 0 \le l < 2^n \\0 \, , & \text{otherwise} \end{cases} \; . \label{eq:X}
\end{equation}
Put in words, this means that the $k$-th bit of each set of $n$ bits
is set to 1, corresponding to the $k$-th image qubit for each domain
qubit.
For example, an $X$ gate on the second of two image
bits would read $\mathcal{C}_{X, 2} = 01010101$.

\subsection{Factorization}

We want to show that an oracle with CNOT- and $X$-gates can indeed
realize any instance $f$ of a periodic (modulo 2) bitwise function. 
For that, we outline here a deterministic procedure to implement the 
oracle for every possible function $f$.
To do this, we will add the characteristics of the possible gates to the
characteristic of $f$ until the resulting string no longer contains non-zero
bits.
The steps required will tell us what combination of gates is
needed to realize $f$.

The function $f$ is represented by its characteristic $\mathcal{C}_f$.
We denote the set of gates that can realize the oracle for any periodic
function as
$S = \{\text{CNOT}_{jk}, X_{k}\}$, where
$1 \le j, k \le n$.
As an example, the set $S$ for two qubits is displayed in Table~\ref{tab:S}.
We need to determine which subset of $S$ will implement a particular
realization of $f$.
This is a factorization of $\mathcal{C}_f$ within the set of
characterizations $\mathcal{C}[S] = \{\mathcal{C}(s)\,; \; s \in S\}$
of the set of gates $S$.

\begin{table}
	\begin{center}
	\begin{tabular}{l|cccccc}
		gate & CNOT$_{11}$ & CNOT$_{12}$ & CNOT$_{21}$ & CNOT$_{22}$ & $X_1$ & $X_2$ \\ \hline
		char. & 00001010 & 00000101 & 00100010 & 00010001 & 10101010 & 01010101 \\
		angle & $\alpha$ & $\beta$ & $\gamma$ & $\delta$ & $\phi$ & $\eta$
	\end{tabular}
	\label{tab:S}
	\caption{The set $S = \{\text{CNOT}_{jk}, X_{k}\} \; (1 \le j,
	k \le n)$, along its characterizations, for two qubits.} 
	\end{center}
\end{table}

This factorization can be carried out in a deterministic, algorithmic
way.
Notice, for example, that in Table~\ref{tab:S}, the first nonzero bit
of each element $\text{CNOT}_{jk}$ is in a different position.
The only operation that will change the first nonzero bit is an $X$ gate
acting on the first image qubit.
Hence, adding the characteristic of said gate to the
characteristic $\mathcal{C}_f$ with a leading 1 (modulo 2) will flip
the first bit to zero.
Note, however, that it will also change other bits.
Thus, we need to step through the bits of $\mathcal{C}_f$ from left to
right.
Each time a non-zero bit appears, we add the characteristic of the gate with
corresponding first non-zero character.
After the application of at most $n 2^{n-1}$ gates, the bit string
will consist of only zeros.
The sequence of gates required to do this uniquely determines the circuit
that realizes our chosen $f$.

\paragraph{Example:} Consider the two-qubit function $f$ with
$f(00), f(01), f(10), f(11) = 10, 11, 11, 10$.
As is easily seen, $f$ has a period $s = 11$ and its characteristic is
$\mathcal{C}_f = 10111110$.
The characteristics of the possible gates can be
read off from Table~\ref{tab:S}.
Following the procedure outlined above, we obtain
\begin{equation}
\mathcal{C}_f = 10111110 \stackrel{X_1}{\longrightarrow}
00010100 \stackrel{\text{CNOT}_{22}}{\longrightarrow}
00000101 \stackrel{\text{CNOT}_{12}}{\longrightarrow} 00000000 \, .
\end{equation}
Thus, the oracle for this particular realization of the 2-bit
periodic function $f$ is implemented by the gate sequence
$\text{CNOT}_{12}\text{CNOT}_{22}X_1$.

\section{Measurement-based quantum computing}
\label{sec:MBQC}

In the preceding section, we have formulated a general, deterministic
algorithm to generate a sequence of gates that realizes a particular
instance of the oracle for the Simon algorithm.
In doing so, we have described both the function and the gates
within the circuit model of quantum computing.
Our task now is to switch to the representation of MBQC, which we will
do using the framework of the ZX-calculus~\cite{ZX-calculus}.
This intuitive and universal graphical calculus was introduced as a
tool for carrying out derivations in multi-qubit quantum computation and
information.
Here we will only give a very brief summary of its relevant features
as well as the specific rules that are needed to carry out the
transition from the circuit to the MBQC formulation.
A somewhat more extensive introduction that is specific to such a
reformulation can be found in Ref.~\citen{3-qubit-DJ-MBQC}.
For a comprehensive overview that both describes the formalism and
illustrates the power of ZX-calculus for a much wider field of
applications, we refer the reader to
Ref.~\citen{Circuit-extraction-tale}.

\subsection{ZX-calculus}
\label{subsec:ZX-calc}
The ZX-calculus is a description of quantum states or quantum
operations in terms of graphs that link two types of nodes:
$Z$-nodes, colored green(or lightly shaded for the visually impaired
reader) here,
and $X$-nodes, colored red (darkly shaded).
On a tensor level, the X-nodes are
defined as
\begin{equation}
	\zx{\leftManyDots{} \zxZ{\alpha} \rightManyDots{}} = \ket{0\dotsc 0}\bra{0\dotsc0} + \mathrm{e}^{\mathrm{i}\alpha} \ket{1\dotsc1}\bra{1\dotsc1} \label{eq:Z-nodes}
\end{equation}
and the Z-nodes as
\begin{equation}
	\zx{\leftManyDots{} \zxX{\alpha} \rightManyDots{}} = \ket{+\dotsm +}\bra{+\dotsm+} + \mathrm{e}^{\mathrm{i}\alpha} \ket{-\dotsm-}\bra{-\dotsm-} \; , \label{eq:X-nodes}
\end{equation}
where $\ket{\pm}=\ket{0}\pm\ket{1}$
(omitting normalization factors).
Incoming and outgoing legs represent degrees of freedom, i.e., tensor
indices.
We note that all ZX-diagrams were typeset with the \emph{zx-calculus}
package \cite{ZX-TeX}.

Connecting nodes via edges leads to diagrams that represent either quantum
states or quantum operators, depending on the configuration of the outgoing edges.
As both states and operators are represented as tensors, we will
not explicitly distinguish between
them; the nature of a diagram should be clear from the context.
As a simple example, the eigenstate of the first Pauli matrix $X$ can be
written as $\zx{\zxZ{} \rar &[\zxwCol] \zxN{}}= \ket{+}$, where we
omit the relative phase angle $\alpha$ of the node when it is zero.

Calculations are carried out in the ZX-calculus by applying rules to
transform diagrams.
Rules can be derived by expressing a diagram in tensor notation and
transforming, typically simplifying, the tensor contractions.
A comprehensive list of rules is given in 
Ref.~\citen{Circuit-extraction-tale}.
A commonly used special simplification is the representation of the
Hadamard gate as a single blue dashed line:
\begin{equation}
 H = \zx{\zxN{} &[\zxwCol] \zxFracZ{\pi}{2} \lar\rar &
 \zxFracX{\pi}{2} \rar & \zxFracZ{\pi}{2} \rar &[\zxwCol]
 \zxN{}} = \zx{\zxN{} \ar[r, blue, dashed] &[2em]
 \zxN{}} \  .\label{eq:rule_Hadamard}
\end{equation}
In this work, we will utilize the following subset of ZX-calculus rules:
\begin{align}
	\begin{ZX}[ampersand replacement=\&]
		\zxN{} \&[\zxwCol]\&[\zxwCol] \zxN{} \\
		\zxN{} \& \zxX{\alpha} \ar[l, 3 vdots] \ar[lu, (] \ar[dl, )] \ar[ur, )] \ar[dr, (] \ar[r, 3 vdots] \& \zxN{} \zxN{} \& \\
		\zxN{} \&\& \zxN{} 
	\end{ZX} &=
	\begin{ZX}[ampersand replacement=\&]
		\zxN{} \&[\zxwCol]\&[\zxwCol] \zxN{} \\
		\zxN{} \& \zxZ{\alpha} \ar[l, 3 vdots] \ar[lu, blue, dashed, (] \ar[dl, blue, dashed, )] \ar[ur, blue, dashed, )] \ar[dr, blue, dashed, (] \ar[r, 3 vdots] \& \zxN{} \\
		\zxN{} \&\& \zxN{}
	\end{ZX} \; , \label{eq:rule_colorswap} \\
	\begin{ZX}[ampersand replacement=\&]
		\leftManyDots{} \zxZ{\alpha} \ar[d, 3 dots]\ar[d, (]\ar[d, )] \rightManyDots{} \\[\zxWRow]
		\leftManyDots{} \zxZ{\beta} \rightManyDots{}
	\end{ZX} &=
	\begin{ZX}[ampersand replacement=\&]
		\leftManyDots{} \zxZ{\alpha+\beta} \rightManyDots{}
	\end{ZX} \; , \label{eq:rule_contraction}\\
	\begin{ZX}[ampersand replacement=\&]
		\zxN{} \&[\zxwCol] \zxZ{} \ar[l, blue, dashed] \ar[r, blue, dashed] \&[\zxwCol] \zxN{}
	\end{ZX} &=
	\begin{ZX}[ampersand replacement=\&]
		\zxN{} \rar \&[\zxWCol] \zxN{}
	\end{ZX} \; , \label{eq:rule_hadamard} \\
	\begin{ZX}[ampersand replacement=\&]
		\zxN{} \& \&[\zxWCol] \zxN{} \\
		\zxX{} \rar \& \zxZ{\alpha} \ar[ur, )] \ar[dr, (] \ar[r, 3 vdots] \& \zxN{} \\
		\zxN{} \& \& \zxN{}
	\end{ZX} &=
	\begin{ZX}[ampersand replacement=\&]
		\zxX{} \rar \&[\zxWCol] \zxN{} \\
		\zxN{} \ar[r, 3 vdots] \& \zxN{} \\
		\zxX{} \rar \& \zxN{}
	\end{ZX} \; . \label{eq:rule_decoupling}
\end{align}
Note that rules (\ref{eq:rule_colorswap})-(\ref{eq:rule_decoupling})
are also valid when Z-nodes and X-nodes are interchanged in the
diagrams. A phenomenological discussion of this set of rules can be found in Ref.~\citen{3-qubit-DJ-MBQC}.
All of these properties can be derived canonically within the
Hilbert-space model by recognizing that a Hadamard gate is a basis
transformation from the Z- to the X-basis and vice versa.

As stated in the preceding section, we need CNOT gates and
$X$ gates in order to implement an any particular realization of the
oracle of the Simon algorithm.
In order to do this in ZX-calculus language,
we note that we can write a CNOT operation on two qubits as 
\begin{equation}
	\begin{quantikz}[align equals at=1.5, thin lines, row sep=0.4cm]
		\qw & \ctrl{1} & \qw \\ \qw & \targ{} & \qw
	\end{quantikz} = 
	\begin{ZX}[row sep=0.4cm]
		\zxN{} &[\zxwCol] \zxZ{} \lar \rar \dar &[\zxwCol]
                \zxN{} \\
                \zxN{} & \zxX{} \lar \rar &[\zxwCol] \zxN{}
	\end{ZX} \; . \label{eq:translation_CNOT}
\end{equation}
An $X$ gate can be written in the ZX-calculus simply as a phase-$\pi$
$X$-node with one input and one output, i.e.,
\begin{equation}
	\begin{quantikz}[align equals at=1, thin lines]
		\qw & \gate{X} & \qw
	\end{quantikz} = 
	\begin{ZX}[align equals at=1]
		\zxN{} &[\zxwCol] \zxX{\pi} \lar \rar &[\zxwCol] \zxN{} \\
	\end{ZX} \; . \label{eq:translation_X}
\end{equation}

Since a general implementation of the oracle requires the ability to either
switch CNOT  and $X$ gates on or off, we utilize an adaptive description
that can control whether or not these gates appear in the oracle.
In particular, we use the adaptive version of a CNOT gate,
\begin{equation}
	\begin{ZX}[align equals at=2]
		\zxN{} \rar &[\zxwCol] \zxZ{} \rar \ar[d, blue, dashed] & \zxN{} \\[\zxHRow]
		\zxN{} & \zxZ{} \ar[d, blue, dashed] \rar & \zxFracZ{\pi}{2} \\[\zxHRow]
		\zxN{} \rar & \zxZ{} \rar & \zxN{}
	\end{ZX} =
	\begin{ZX}[align equals at=2]
		\zxN{} \rar &[\zxwCol] \zxZ{} \rar \ar[d, blue, dashed] & \zxN{} \\[\zxHRow]
		\zxN{} & \zxFracZ{\pi}{2} \ar[d, blue, dashed] & \\[\zxHRow]
		\zxN{} \rar & \zxZ{} \rar & \zxN{}
	\end{ZX} =
	\begin{ZX}[align equals at=1.5]
		\zxN{} \rar &[\zxwCol] \zxFracZ-{\pi}{2} \rar \ar[d, blue, dashed]&[\zxwCol] \zxN{} \\[0.55cm]
		\zxN{} \rar & \zxFracZ-{\pi}{2} \rar & \zxN{}
	\end{ZX} =
	\begin{ZX}[align equals at=1.5]
		\zxN{} \rar &[\zxwCol] \zxZ{} \rar \ar[d, blue, dashed] & \zxFracZ-{\pi}{2} \rar &[\zxwCol] \zxN{} \\[0.55cm]
		\zxN{} \rar & \zxZ{} \rar & \zxFracZ-{\pi}{2} \rar & \zxN{}
	\end{ZX} =
	\begin{ZX}[align equals at=1.5]
		\zxN{} \rar &[\zxwCol] \zxZ{} \rar \dar & \zxFracZ-{\pi}{2} \rar &[\zxwCol] \zxN{} \\[0.55cm]
		\zxN{} \ar[r, blue, dashed] & \zxX{} \ar[r, blue, dashed] & \zxFracZ-{\pi}{2} \rar & \zxN{}
	\end{ZX}
	\label{eq:adaptive_CNOT}
\end{equation}
\begin{equation}
	\begin{ZX}[align equals at=2]
		\zxN{} \rar &[\zxwCol] \zxZ{} \rar \ar[d, blue, dashed] &[\zxwCol] \zxN{} \\[\zxHRow]
		\zxN{} & \zxZ{} \ar[d, blue, dashed] \rar & \zxX{} \\[\zxHRow]
		\zxN{} \rar & \zxZ{} \rar & \zxN{} 
	\end{ZX} =
	\begin{ZX}[align equals at=2.5]
		\zxN{} \rar &[\zxwCol] \zxZ{} \rar \ar[d, blue, dashed] &[\zxwCol] \zxN{} \\[\zxHRow]
		\zxN{} & \zxX{} & \zxN{} \\
		\zxN{} & \zxX{} \ar[d, blue, dashed] & \zxN{} \\[\zxHRow]
		\zxN{} \rar & \zxZ{} \rar & \zxN{} 
	\end{ZX} =
	\begin{ZX}[align equals at=2.5]
		\zxN{} \rar &[\zxwCol] \zxZ{} \rar \dar &[\zxwCol] \zxN{} \\
		\zxN{} & \zxZ{} & \zxN{} \\
		\zxN{} & \zxZ{} \dar & \zxN{} \\
		\zxN{} \rar & \zxZ{} \rar & \zxN{} 
	\end{ZX} =
	\begin{ZX}[align equals at=1.5]
		\zxN{} \rar &[\zxwCol] \zxZ{} \rar&[\zxwCol] \zxN{} \\[0.5cm]
		\zxN{} \rar & \zxZ{} \rar & \zxN{} 
	\end{ZX} \; .
	\label{eq:adaptive_bridge}
\end{equation}
Eq.~(\ref{eq:adaptive_CNOT}) tells us that a configuration
with a terminal $\pi/2$-Z-node that is attached to a phase-zero
Z-node, which is, in turn, 
connected to two further Z-nodes via Hadamard-edges is equivalent to the CNOT
gate introduced in eq.~(\ref{eq:CNOT}), aside from some phase
corrections on the neighborhood.
These required correcting phases take on the form of two
$\pi/2$-Z-nodes and two Hadamard nodes and need to be included
when this ZX-network fragment is embedded in a larger ZX-diagram.
Note that the protruding $\pi/2$-node in the leftmost part of the
equation represents the bra $\bra{0} + \mathrm{i}\bra{1}$ and thus acts 
as a measurement of the unit-eigenvalue eigenstate of the $Y$-Pauli
matrix.
We will treat measurements in the scope of the ZX-calculus more
thoroughly in the next section.

Eq.~(\ref{eq:adaptive_bridge}) shows that inserting a zero-phase
$X$-node rather than $\pi/2$-Z-node into eq.~\eqref{eq:adaptive_CNOT}
disconnects the two $Z$-nodes on the top and on the bottom of
the diagram.
Thus, one  can control whether or not a CNOT gate between
two nodes is present by choosing whether there is a $\pi/2$-$Z$-node
or a zero-phase $X$-node.
Note that the identities \eqref{eq:adaptive_CNOT}
and \eqref{eq:adaptive_bridge} can be derived by applying
rules \eqref{eq:rule_hadamard}-\eqref{eq:rule_decoupling}.

\subsection{MBQC with the ZX-calculus}
\label{subsec:MPQC_ZX}
For the purposes of this work, the essential feature of
a ZX-diagram-based picture is that it provides a natural way to describe
measurement-based quantum computing.
In order to see why this is, we realize that a ZX-diagram with only Z-nodes
(green) and in which all edges are Hadamard edges
(blue-dashed) represents a \emph{cluster state} in which
the Z-nodes represent the qubits \cite{Circuit-extraction-tale}.
A projective measurement with outcome 0 of a qubit along an
axis in the $x$-$y$-plane
can be written as the projector
$\zx{\zxN{} &[\zxWCol] \zxZ{\alpha} \lar}$, where $\alpha$ is the
angle of the basis in the plane.
For example, a measurement along
the $x$-axis with outcome
0, corresponding
to the projected state $\ket{+}=\ket{0}+\ket{1}$, would be
represented by $\zx{\zxN{} &[\zxWCol] \zxZ{} \lar}$.
Since all nodes in a cluster state are Z-nodes, one can contract this
measurement projection to the associated qubit it is attached to using
rule (\ref{eq:rule_contraction}).
Hence, a ZX-diagram consisting of only Z-nodes and Hadamard edges
can also be understood as a sequence of $x$-measurements on a cluster
state.
On a technical note,
for MBQC within ZX-calculus, it is usually assumed for simplicity
that any
measurement always has the
outcome 0.
If the other outcome, 1, were to occur instead,
this ``error'' could be propagated
through the network so that either
subsequent measurement axes need to
be adjusted or the final measurement outcome has to be altered in a manner
that directly follows from the phases that occur.
We will not elucidate this procedure further here;
see Ref.~\citen{Circuit-extraction-tale} for a detailed explanation.

\subsection{Translation to MBQC}
\label{subsec:translationMBQC}

In Secs.~\ref{subsec:ZX-calc} and \ref{subsec:MPQC_ZX}, we
introduced the prerequisites for characterizing MBQC with the
ZX-calculus and described several identities that are required to translate the
Simon algorithm to a measurement-based form.
Here we will first develop a procedure for the two-qubit algorithm
and then explain how it can be expanded to the $n$-qubit case.

Fig.~\ref{fig:general-circuit} depicts a schematic circuit for the
two-qubit Simon algorithm that includes all gates that could occur in
a specific realization of the oracle: CNOT-gates between
all pairs of input and auxiliary qubits and $X$-gates on all auxiliary qubits.
Some (or all) of these gates may not be present in a circuit realizing
a particular instance of the oracle.
Thus, an arbitrary oracle can be constructed by switching on or off
particular gates.
A general circuit for which each gate can be switched on
or switched off can be realized in the ZX-calculus using the
adaptive description of the CNOT gate in Eq.~\eqref{eq:adaptive_CNOT}
and Eq.~\eqref{eq:adaptive_bridge}, along with the rules in
Eqs.~\eqref{eq:rule_hadamard}-\eqref{eq:rule_decoupling}.

\begin{figure}
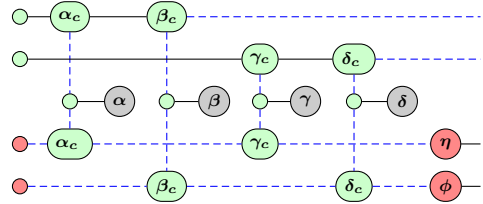

\begin{center}
	\begin{ZX}
		\zxZ{} \rar &[\zxHCol] \zxZ{\alpha_c} \ar[rr] \ar[dd, blue, dashed] && \zxZ{\beta_c} \ar[rrrrrrr, blue, dashed] \ar[dd, blue, dashed] &&&&&&&[\zxwCol] \zxN{} \\
		\zxZ{} \ar[rrrrr] &&&&& \zxZ{\gamma_c} \ar[rr] \ar[d, blue, dashed] && \zxZ{\delta_c} \ar[rrr, blue, dashed] \ar[d, blue, dashed] &&& \zxN{} \\
		& \zxZ{} \rar \ar[d, blue, dashed] & \zxZ[fill=black!20]{\alpha} & \zxZ{} \ar[dd, blue, dashed] \rar & \zxZ[fill=black!20]{\beta} & \zxZ{} \ar[d, blue, dashed] \rar & \zxZ[fill=black!20]{\gamma} & \zxZ{} \ar[dd, blue, dashed] \rar & \zxZ[fill=black!20]{\delta} && \\
		\zxX{} \ar[r, blue, dashed] & \zxZ{\alpha_c} \ar[rr, blue, dashed] && \zxN{} \ar[rr, blue, dashed] && \zxZ{\gamma_c} \ar[rrrr, blue, dashed] &&&& \zxX{\eta} \rar & \zxN{} \\
		\zxX{} \ar[rrr, blue, dashed] &&& \zxZ{\beta_c} \ar[rr, blue, dashed] && \zxN{} \ar[rr, blue, dashed] && \zxZ{\delta_c} \ar[rr, blue, dashed] && \zxX{\phi} \rar & \zxN{}
	\end{ZX}
\end{center}
\caption{Raw translation of the two-qubit Simon algorithm into a ZX-diagram.}
\label{fig:raw-translation}
\end{figure}

A raw translation of the two-qubit circuit is depicted in
Fig.~\ref{fig:raw-translation}.
Note that we have identified the initial states of the qubits as
$\ket{+} = \zx{\zxZ{} \rar &[\zxwCol] \zxN{}}$ and
$\ket{0} = \zx{\zxX{} \rar &[\zxwCol] \zxN{}}$.
The adaptive CNOT gates are controlled by a
node that we depict as
\begin{equation}
	\begin{ZX}
		\zxZ[fill=black!20]{\alpha} \rar &[\zxwCol] \zxN{}
	\end{ZX} = \begin{cases} \begin{ZX}
		\zxX{} \rar &[\zxwCol] \zxN{}
	\end{ZX} \, , & \alpha = 0 \\ \begin{ZX}
		\zxFracZ{\pi}{2} \rar &[\zxwCol] \zxN{}
	\end{ZX} \; , & \alpha = \frac{\pi}{2} \end{cases} \; .
	\label{eq:adaptive_angles}
\end{equation}
This node is adaptive in the sense that it is chosen to be
either an $X$-measurement or a $Y$-measurment depending on the
particular realization of the function that the oracle represents.
Which of the two permitted distinct phases, $\alpha = 0$
or $\alpha = \pi/2$, is chosen determines the color of the node.
As expressed in eq.~\eqref{eq:adaptive_CNOT}, the adaptive CNOT gate
also requires some extra correcting phases to be applied to the two
qubits adjacent to the one being measured.
Those are included into the raw translation by
\begin{equation}
	\alpha_c = \begin{cases} 0 & \alpha = 0 \\ \frac{\pi}{2} & \alpha = \frac{\pi}{2} \end{cases} \; ,
	\label{eq:corrective_angles}
\end{equation}
where $\alpha$ is the same value as in eq.~\eqref{eq:adaptive_angles}.
The same scheme applies to all adaptive angles.
The nodes in Fig.~\ref{fig:raw-translation} with angles $\eta$ and
$\phi$ represent the $X$ gates; the phase of $\eta$ or
$\phi$ is $\pi$ if an $X$ gate is present and $0$ if not.

Starting from the raw translation of
Fig.~\ref{fig:raw-translation}, we rewrite and transform
the graph into a form consistent with MBQC using the rules presented in
Eqs.~\eqref{eq:rule_hadamard}-\eqref{eq:rule_decoupling},
yielding Fig.~\ref{fig:MBQC-simon}.
A detailed description of all individual steps taken is given in
the appendix to this work; here we give an overview and describe the
most important steps.

First, notice that the two protruding edges on the bottom in
Fig.~\ref{fig:raw-translation} represent the auxiliary qubits.
Their final state is irrelevant because only the first two (working)
qubits are measured.
In fact, anything that is done to these branches of the diagram is irrelevant;
we can simply omit the bottom legs to simplify the diagram.

A series of relatively straightforward applications of the rules of
ZX-diagrams then leads to the diagram depicted in
Fig.~\ref{fig:MBQC-simon}, which implements the two-qubit Simon
algorithm in the framework of MBQC.
As is required for MBQC, it can be seen that all qubits of the
many-body system are $Z$-nodes.
Measurements on these qubits are always carried out in the
$y$-$z$-plane, at an angle that depends on the corrective factors.
The four remaining qubits will be measured in the adaptive manner
described above.
They are depicted as grey nodes, which either represent a
$\pi/2$-$z$-measurement or a zero-phase $x$-measurement.

\begin{figure}
\begin{center}
	\begin{ZX}
		\zxZ[fill=black!20]{\alpha} \dar & \zxZ{\alpha_c+\beta_c} \ar[dl, (, blue, dashed] \ar[r, blue, dashed] \ar[d, blue, dashed] & \zxZ{} \rar &[\zxwCol] \zxN{} \\
		\zxZ{} \ar[d, blue, dashed] & \zxZ{} \ar[d, blue, dashed] \rar & \zxZ[fill=black!20]{\beta} & \\
		\zxZ{\alpha_c + \gamma_c + \eta} \ar[d, blue, dashed] & \zxZ{\beta_c + \delta_c + \phi} \ar[d, blue, dashed] & & \\
		\zxZ{} \dar \ar[dr, (, blue, dashed] & \zxZ{} \ar[d, blue, dashed] \rar & \zxZ[fill=black!20]{\delta} & \\
		\zxZ[fill=black!20]{\gamma} & \zxZ{\gamma_c + \delta_c} \ar[r, blue, dashed] & \zxZ{} \rar & \zxN{}
	\end{ZX}
\caption{Measurement-based form of the 2-qubit Simon algorithm.}
\label{fig:MBQC-simon}
\end{center}
\end{figure}

To clarify the meaning of the MBQC protocol in terms of the ZX-calculus,
we give here a short explanation of how to interpret
Fig.~\ref{fig:MBQC-simon}.
First, a cluster state, consisting of qubits ordered in a graph as
the green nodes connected by blue and dashed lines, is created.
Then, all qubits of the cluster state except for the two green nodes to the
right are measured according to the angles depicted in the figure.
If outcomes other than 0 occur during a measurement, corrective
factors must be propagated through the graph; subsequent
measurement angles and/or the final result might have to be adapted.
Eventually, the two qubits on the right are measured in the $x$-basis.
For an oracle representing a periodic function, the bit string $m$
that is measured satisfies the condition $m \cdot s = 0$, where $s$ is
the period.
Finally, it is necessary to repeat the computation sufficiently often to
obtain $n$ linearly independent equations for $s$ (where $n=2$ here),
and these equations must be solved classically, just as in the
quantum-circuit formulation of the algorithm.

\section{$n$-qubit Simon algorithm}
\label{sec:nbit}

In Sec.~\ref{subsec:translationMBQC}, we derived the MBQC formulation of the
two-qubit Simon algorithm.
Here we will expand this formulation to treat $n$ qubits.
The translation scheme and the following measurement procedure is essentially
the same as in the two-qubit case.
In the following, we will construct the graphs of the cluster states that are required to
formulate the $n$-qubit Simon algorithm in terms of MBQC.

In the preceding section, we have learned that a general instance of
the oracle of the Simon algorithm can be constructed using a network
consisting of CNOT gates between arbitrary unique pairs of working and
auxiliary qubits.
In total, there are $n^2$ possible CNOT gates, each of which can be
realized by an adaptive CNOT gate in the ZX-language.
In the ZX-scheme, each pair of working and auxiliary qubits can be pulled together
to form one node in the ZX-diagram, so that we obtain $n^2$ adaptive CNOT
gates connecting these nodes.
In the two-qubit case, depicted in Fig.~\ref{fig:MBQC-simon}, we have
four qubits, connected by $2^2 = 4$ adaptive CNOT gates.
Each node is connected to exactly $n$ other gates.

In order to better display the topology of the graphs of $n$-qubit,
measurement-based oracles, we introduce a compact notation for an adaptive
CNOT gate.
We define red and dashed lines as 
\begin{equation}
	\zx{\zxN{} \ar[r, red, dashed] &[2em] \zxN{}} \equiv \zx[align equals at=2]{\zxN{} &[1em] \zxZ[fill=black!20]{\cdot} \dar &[1em] \zxN{} \\ \zxN{} \ar[r, blue, dashed] & \zxZ{} \ar[r, blue, dashed] & \zxN{}} \; , 
\end{equation}
where the dot in the grey node represents the measurement defined in
eq.~(\ref{eq:adaptive_angles}) in which the choice of measurement
angle determines the presence or absence of a CNOT gate.
Since we do not specify the measurement angle explicitly in this notation, we
also omit the corrective angles by writing
\begin{equation}
	\zx{\leftManyDots{} \zxZ{\alpha_c + \beta_b + \dotsm} \rightManyDots{}} = \zx{\leftManyDots{} \zxZ{\cdot} \rightManyDots{}} \; , 
\end{equation}
where the corrective angles to be applied to the qubits neighboring
the measured node are specified by eq.~\eqref{eq:corrective_angles}.

The form of the graph for the cluster state of the $n$-qubit
version of the Simon algorithm is determined by the number $n$ of working
and auxiliary qubits as well as their connections through adaptive
CNOT gates.
For visualization, we arrange all qubits on a ring, alternating between
auxiliary and working qubits.
We then connect each working qubit with all auxiliary qubits via red
and dashed lines, which represent adaptive CNOT gates.
The working qubits are distinguished by protruding legs that are used for
the final measurement.
Thus, the $n$-qubit algorithm is represented by a ring-like graph
consisting of $2n$ nodes, $n^2$ edges, and $n$ protruding legs.

\begin{figure}\begin{center}
	\begin{tikzpicture}[znode/.style={draw, fill=green!20!white, font={\fontsize{8}{10}\selectfont\boldmath}, inner sep=2pt, outer sep=0pt, rounded rectangle}, baseline=(base)]
		\coordinate (base) at (0, 0);
		\node[znode] (1) at (1, 0) {$\cdot$};
		\node[znode] (2) at (0, 1) {$\cdot$};
		\node[znode] (3) at (-1, 0) {$\cdot$};
		\node[znode] (4) at (0, -1) {$\cdot$};
		\draw[red, dashed] (1) -- (2) -- (3) -- (4) -- (1);
		\draw (2) -- ++(0, 0.5);
		\draw (4) -- ++(0, -0.5);
		\node at (-1.5, 1.5) {(a)};
	\end{tikzpicture}
	\hspace{2em}
	\begin{tikzpicture}[znode/.style={draw, fill=green!20!white, font={\fontsize{8}{10}\selectfont\boldmath}, inner sep=2pt, outer sep=0pt, rounded rectangle}, baseline=(base)]
		\coordinate (base) at (0, 0);
		\node[znode] (1) at (0.87, 0.5) {$\cdot$};
		\node[znode] (2) at (0, 1) {$\cdot$};
		\node[znode] (3) at (-0.87, 0.5) {$\cdot$};
		\node[znode] (4) at (-0.87, -0.5) {$\cdot$};
		\node[znode] (5) at (0, -1) {$\cdot$};
		\node[znode] (6) at (0.87, -0.5) {$\cdot$};
		\draw[red, dashed] (1) -- (2) -- (3) -- (4) -- (5) -- (6) -- (1) -- (4);
		\draw[red, dashed] (2) -- (5);
		\draw[red, dashed] (3) -- (6);
		\draw (2) -- ++(0, 0.5);
		\draw (4) -- ++(-0.44, -0.25);
		\draw (6) -- ++(0.44, -0.25);
		\node at (-1.5, 1.5) {(b)};
	\end{tikzpicture}
	\hspace{2em}
	\begin{tikzpicture}[znode/.style={draw, fill=green!20!white, font={\fontsize{8}{10}\selectfont\boldmath}, inner sep=2pt, outer sep=0pt, rounded rectangle}, baseline=(base)]
		\coordinate (base) at (0, 0);
		\node[znode] (1) at (1, 0) {$\cdot$};
		\node[znode] (2) at (0.71, 0.71) {$\cdot$};
		\node[znode] (3) at (0, 1) {$\cdot$};
		\node[znode] (4) at (-0.71, 0.71) {$\cdot$};
		\node[znode] (5) at (-1, 0) {$\cdot$};
		\node[znode] (6) at (-0.71, -0.71) {$\cdot$};
		\node[znode] (7) at (0, -1) {$\cdot$};
		\node[znode] (8) at (0.71, -0.71) {$\cdot$};
		\draw[red, dashed] (1) -- (2) -- (3) -- (4) -- (5) -- (6) -- (7) -- (8) -- (1) -- (4) -- (7) -- (2) -- (5) -- (8) -- (3) -- (6) -- (1);
		\draw (1) -- ++(0.5, 0);
		\draw (3) -- ++(0, 0.5);
		\draw (5) -- ++(-0.5, 0);
		\draw (7) -- ++(0, -0.5);
		\node at (-1.5, 1.5) {(c)};
	\end{tikzpicture}
	\caption{Topology of the cluster state required to carry
		out the (a) two-, (b) three- and (c) four-qubit
		Simon-algorithm in MBQC.}
	\label{fig:schematic}
\end{center}\end{figure}
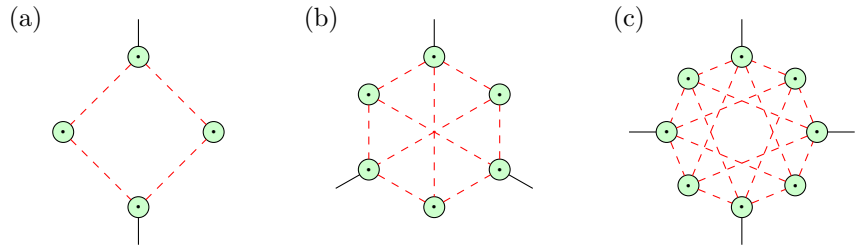

As an example, recall the pattern of the oracle for the
two-qubit algorithm, Fig.~\ref{fig:MBQC-simon}.
We place the four qubits (2 working and 2 auxiliary) on a ring,
forming a square.
Connecting working to auxiliary qubits and introducing protruding
legs leads to the schematic diagram depicted in Fig.~\ref{fig:schematic}(a).
The scheme is easily extended to three qubits,
Fig.~\ref{fig:schematic}(b), and four qubits, Fig.~\ref{fig:schematic}(c).

\section{Summary and Discussion}
\label{sec:conclusion}

In this work, we have reformulated the oracles of the Simon
period-finding algorithm in the framework of MBQC, translating quantum-circuit
descriptions to ZX-diagrams and then bringing the ZX-diagrams into a form that
represents a measurement-based algorithm on a cluster state.
An essential intermediate step is the construction of a deterministic algorithm
to represent any given periodic function as a quantum circuit
consisting only of CNOT and $X$ gates.
We have constructed such a MBQC-suited ZX-network explicitly for the
two-qubit Simon algorithm; the resulting cluster state consists of
eight qubits ordered in the shape of a square. 
Furthermore, we have described how to construct cluster states
for the general $n$-qubit algorithm and have given explicit forms for
such states in the three- and four-qubit cases.

The deterministic algorithm to transform any periodic function into
a circuit representation using CNOT and $X$ gates that we have developed
in Sec.~\ref{sec:simon} is an essential part of the procedure to
reformulate the Simon algorithm in MBQC.
We know of no such constructive quantum-circuit-based algorithm in the
literature. 
Algorithms to construct an oracle for the Simon algorithm, such as
that formulated in Ref.~\citen{Simon-Qiskit}, do exist.
However, in our understanding, they are all based on the
assumption that a specific fixed period $s$ is given.
In our opinion, assuming a given period in order to construct an
oracle is somewhat circular because the whole purpose of the
calculation is to determine an unknown period.
In our algorithm, such an assumption
would strongly limit the allowed
inputs and outputs of the function.
It can be argued that such a restriction also, essentially, contains
knowledge of the period.
Finding the period of a function with no prior knowledge is, in out opinion, the 
point of applying the Simon algorithm.
We remark, however, that 
guessing the period and subsequently constructing the
algorithm around it will become exponentially difficult as the
function become larger.
We appreciate that the question of how to design oracles for oracular
quantum algorithms and, in particular, the issue of how much presupposed
knowledge about the function or oracle is available,
is still a subject of discussion and, at this point
in time, is a somewhat philosophical issue.

Specific quantum circuits for the oracle of the two-qubit Simon algorithm have been 
published in Ref.~\citen{Experimental-Simon}, albeit without
specifying a rigorous algorithm to construct them.
We comment that the two-qubit case is of sufficiently low complexity
that a working, and presumably optimal, MBQC implementation can be
found relatively easily by educated guessing; in our opinion, this is
no longer possible for larger systems.
Our approach of deterministically constructing implementations of a
general oracle for a given algorithm is in line with other published work
such as Ref.~\citen{4qubit-DJ}, in which arbitrary oracle circuits are
constructed for the four-qubit Deutsch-Jozsa algorithm.

The topology of the cluster states that we have constructed to implement
a general $n$-qubit Simon algorithm gives information about the
complexity of the algorithm.
We have shown that our graph for the general algorithm consists of $2n$
nodes connected by $n^2$ edges.
Each of these generalized edges consists of one measurement node and two basic
cluster-state edges.
Thus, as expected, the quantum resources required for the algorithm grow only
polynomially with the size of the problem.
On the other hand, only the two-qubit Simon algorithm
can be realized as a planar (two-dimensional) cluster state.
For all larger variants, the graphs do have two-dimensional cross
sections, but require a higher number of dimensions for their
implementation.

This topological complexity is especially relevant for experimental
implementation.
A simplified version of the two-bit Simon algorithm has already been
experimentally implemented by the authors of Ref.~\citen{Experimental-Simon}.
For larger systems, it will be a challenge to implement the cluster states
that we have derived.
Large quantum states on arbitrary graphs are especially difficult to
produce and control, both in solid-state and in photonic systems. 
We have shown that the number of two-dimensional cross sections
increases as the number of bits in the oracle function increase.
The increased interconnectivity between qubits will also make the creation
of the associated cluster states less tractable.

In this work, we have shown that the MBQC graphs for the oracles are equivalent
to their counterparts in the framework of circuit quantum computing.
However, we have neither argued nor proven that these are the cluster
states that require the smallest number of qubits and/or the smallest
interconnectivity.
To do this raises a question of optimal minimization that cannot be
answered through the framework of ZX-calculus alone because ZX-calculus
is only a descriptive language.
Nevertheless, the rules of ZX-calculus are a good tool to visualize a
path to a minimal solution, as is discussed
in Ref.~\citen{ZX-simplification}.
Thus, interesting subjects for future work would be 
to investigate if the patterns derived here do, in fact, yield a minimal
cluster state and, in addition,
to formulate schemes for finding such
patterns in general and for proving that they are, in fact, minimal.

Historically, Simon's formulation of the Simon algorithm
led to Shor's development of his famous
prime-factorization algorithm \cite{Simon, Simon-Interview, Shor}.
This development sequence can be taken as inspiration for the
development of the MBQC-variants of these algorithms.
Since the underlying structure of the hidden subgroup problem in the Simon
algorithm is also an integral part of the Shor algorithm, it seems only natural
to apply the principles of this work to the Shor algorithm.
The methods developed here might additionally be useful for
reformulating other algorithms, information protocols, error correction
codes, etc.\ in a MBQC picture.

\appendix*

\section{Simplification of the oracle in the ZX-language}
\label{app:simplification}

In this appendix, we describe the detailed
steps used to reformulate the raw ZX-translation of the two-qubit
Simon algorithm as a ZX-network suited for MBQC.
For readability, we replicate here
the raw translation of the Simon algorithm in Fig.~\ref{fig:raw-translation},
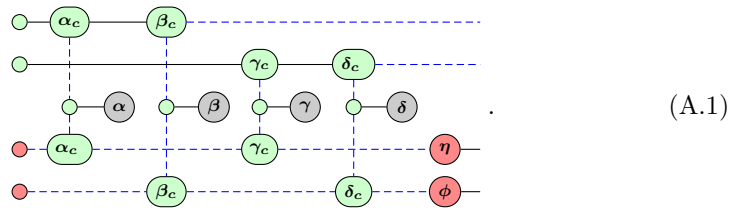
\begin{equation}
 	\begin{ZX}
 		\zxZ{} \rar &[\zxHCol] \zxZ{\alpha_c} \ar[rr] \ar[dd, blue, dashed] && \zxZ{\beta_c} \ar[rrrrrrr, blue, dashed] \ar[dd, blue, dashed] &&&&&&&[\zxwCol] \zxN{} \\
 		\zxZ{} \ar[rrrrr] &&&&& \zxZ{\gamma_c} \ar[rr] \ar[d, blue, dashed] && \zxZ{\delta_c} \ar[rrr, blue, dashed] \ar[d, blue, dashed] &&& \zxN{} \\
 		& \zxZ{} \rar \ar[d, blue, dashed] & \zxZ[fill=black!20]{\alpha} & \zxZ{} \ar[dd, blue, dashed] \rar & \zxZ[fill=black!20]{\beta} & \zxZ{} \ar[d, blue, dashed] \rar & \zxZ[fill=black!20]{\gamma} & \zxZ{} \ar[dd, blue, dashed] \rar & \zxZ[fill=black!20]{\delta} && \\
 		\zxX{} \ar[r, blue, dashed] & \zxZ{\alpha_c} \ar[rr, blue, dashed] && \zxN{} \ar[rr, blue, dashed] && \zxZ{\gamma_c} \ar[rrrr, blue, dashed] &&&& \zxX{\eta} \rar & \zxN{} \\
 		\zxX{} \ar[rrr, blue, dashed] &&& \zxZ{\beta_c} \ar[rr, blue, dashed] && \zxN{} \ar[rr, blue, dashed] && \zxZ{\delta_c} \ar[rr, blue, dashed] && \zxX{\phi} \rar & \zxN{}
 	\end{ZX} \; .
\end{equation}
We apply the identities
\begin{equation}
	\begin{ZX}[ampersand replacement=\&]
		\zxN{} \&[\zxwCol]\&[\zxwCol] \zxN{} \\
		\zxN{} \& \zxX{\alpha} \ar[l, 3 vdots] \ar[lu, (] \ar[dl, )] \ar[ur, )] \ar[dr, (] \ar[r, 3 vdots] \& \zxN{} \zxN{} \& \\
		\zxN{} \&\& \zxN{} 
	\end{ZX} =
	\begin{ZX}[ampersand replacement=\&]
		\zxN{} \&[\zxwCol]\&[\zxwCol] \zxN{} \\
		\zxN{} \& \zxZ{\alpha} \ar[l, 3 vdots] \ar[lu, blue, dashed, (] \ar[dl, blue, dashed, )] \ar[ur, blue, dashed, )] \ar[dr, blue, dashed, (] \ar[r, 3 vdots] \& \zxN{} \\
		\zxN{} \&\& \zxN{}
	\end{ZX} \quad\text{and}\quad
	\begin{ZX}[ampersand replacement=\&]
		\leftManyDots{} \zxZ{\alpha} \ar[d, 3 dots]\ar[d, (]\ar[d, )] \rightManyDots{} \\[\zxWRow]
		\leftManyDots{} \zxZ{\beta} \rightManyDots{}
	\end{ZX} =
	\begin{ZX}[ampersand replacement=\&]
		\leftManyDots{} \zxZ{\alpha+\beta} \rightManyDots{}
	\end{ZX} \; ,
	\label{eq:app_rules_1}
\end{equation}
which are rules~\eqref{eq:rule_colorswap} and \eqref{eq:rule_contraction} in the main text, to obtain an initial simplified version
of the oracle, 
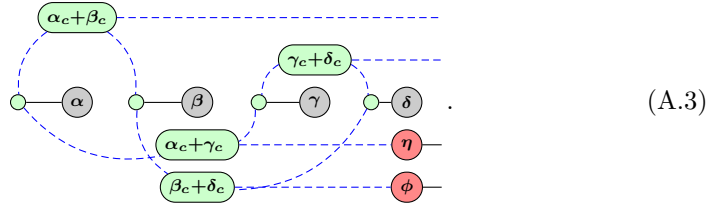
\begin{equation}
	\begin{ZX}
		\zxN{} & \zxZ{\alpha_c+\beta_c} \ar[rdd, ), blue, dashed] \ar[ldd, (, blue, dashed] \ar[rrrrrrr, blue, dashed] &&&&&&&[\zxwCol] \zxN{} \\
		&&&&& \zxZ{\gamma_c + \delta_c} \ar[ld, (, blue, dashed] \ar[rd, ), blue, dashed] \ar[rrr, blue, dashed] &&& \zxN{} \\
		\zxZ{} \rar \ar[drrr, (, blue, dashed] & \zxZ[fill=black!20]{\alpha} & \zxZ{} \ar[ddr, (, blue, dashed] \rar & \zxZ[fill=black!20]{\beta} & \zxZ{} \ar[dl, ), blue, dashed] \rar & \zxZ[fill=black!20]{\gamma} & \zxZ{} \ar[ddlll, ), blue, dashed] \rar & \zxZ[fill=black!20]{\delta} & \\
		&&& \zxZ{\alpha_c + \gamma_c} \ar[rrrr, blue, dashed] &&&& \zxX{\eta} \rar & \zxN{} \\
		&&& \zxZ{\beta_c + \delta_c} \ar[rrrr, blue, dashed]
		&&&& \zxX{\phi} \rar & \zxN{}
	\end{ZX} \, .
\end{equation}
Since we no longer utilize the information encoded in the  auxiliary
qubit after applying the oracle, we can omit the protruding legs to
the right of the red nodes.
More formally, we add a projective measurement
$\zx{\zxN{} &[\zxwCol] \zxX{}\lar}$ in the $x$-basis to annihilate the protruding
legs.
Using  rules \eqref{eq:app_rules_1}, we can contract the legs into
the neighboring Z-nodes.
On the two top-most protruding legs, we then apply the identity
$\zx{\zxN{} \rar &[\zxwCol] \zxZ{} \rar &[\zxwCol] \zxN{}}$
so that they become
regular rather than Hadamard legs.
Hence, we obtain
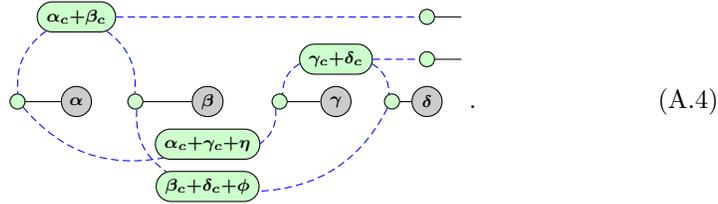
\begin{equation}
	\begin{ZX}
		\zxN{} & \zxZ{\alpha_c+\beta_c} \ar[rdd, ), blue, dashed] \ar[ldd, (, blue, dashed] \ar[rrrrrr, blue, dashed] &&&&&&\zxZ{} \rar &[\zxwCol] \zxN{} \\
		&&&&& \zxZ{\gamma_c + \delta_c} \ar[ld, (, blue, dashed] \ar[rd, ), blue, dashed] \ar[rr, blue, dashed] && \zxZ{} \rar & \zxN{}\\
		\zxZ{} \rar \ar[drrr, (, blue, dashed] & \zxZ[fill=black!20]{\alpha} & \zxZ{} \ar[ddr, (, blue, dashed] \rar & \zxZ[fill=black!20]{\beta} & \zxZ{} \ar[dl, ), blue, dashed] \rar & \zxZ[fill=black!20]{\gamma} & \zxZ{} \ar[ddlll, ), blue, dashed] \rar & \zxZ[fill=black!20]{\delta} & \\
		&&& \zxZ{\alpha_c + \gamma_c + \eta} &&&&&\\
		&&& \zxZ{\beta_c + \delta_c + \phi} &&&&&
	\end{ZX} \; .
\end{equation}
A straightforward reordering of the graph yields the result
 
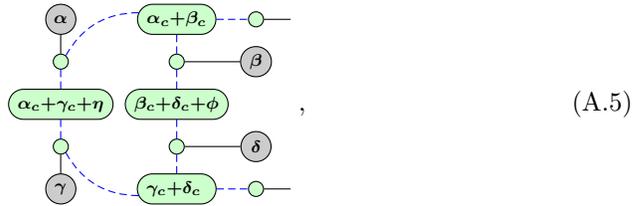
\begin{equation}
 	\begin{ZX}
 		\zxZ[fill=black!20]{\alpha} \dar & \zxZ{\alpha_c+\beta_c} \ar[dl, (, blue, dashed] \ar[r, blue, dashed] \ar[d, blue, dashed] & \zxZ{} \rar &[\zxwCol] \zxN{} \\
 		\zxZ{} \ar[d, blue, dashed] & \zxZ{} \ar[d, blue, dashed] \rar & \zxZ[fill=black!20]{\beta} & \\
 		\zxZ{\alpha_c + \gamma_c + \eta} \ar[d, blue, dashed] & \zxZ{\beta_c + \delta_c + \phi} \ar[d, blue, dashed] & & \\
 		\zxZ{} \dar \ar[dr, (, blue, dashed] & \zxZ{} \ar[d, blue, dashed] \rar & \zxZ[fill=black!20]{\delta} & \\
 		\zxZ[fill=black!20]{\gamma} & \zxZ{\gamma_c + \delta_c} \ar[r, blue, dashed] & \zxZ{} \rar & \zxN{}
 	\end{ZX} \; , 
 \end{equation}
which we also depict in
the main text as Fig.~\ref{fig:MBQC-simon} for readability.


\end{document}